\documentclass[12pt,preprint]{aastex}

\slugcomment{Submit to ApJ}

\shorttitle{The Origin of The EGB}
\shortauthors{Zhou et al.}

\begin{document}


\title{A NEW RESULT ON THE ORIGIN OF THE EXTRAGALACTIC GAMMA-RAY BACKGROUND}


\author{
Ming Zhou\altaffilmark{1,2,3},
Jiancheng Wang\altaffilmark{1,2}
}
\altaffiltext{1} {National Astronomical Observatories, Yunnan Observatory, Chinese Academy
of Sciences,  Kunming 650011, China}
\altaffiltext{2} {Key Laboratory for the Structure and Evolution of Celestial Objects,
Chinese Academy of Sciences,  Kunming 650011, China}
\altaffiltext{3} {Graduate School, Chinese Academy of Sciences, Beijing, P.R. China}

\email{mzhou@ynao.ac.cn}



\begin{abstract}
In the paper, we continually use the method of image stacking to study
the origin of the extragalactic gamma-ray background (EGB) at GeV bands,
and find that the Faint Images of the Radio Sky at Twenty centimeters (FIRST) sources
undetected by the Large Area Telescope
on the {\it Fermi Gamma-ray Space Telescope}
can contribute about (56$\pm$6)\,\% of the EGB. Because the FIRST is a flux
limited sample of radio sources with incompleteness at the faint limit,
we consider that the point-sources, including blazars, non-blazar AGNs,
starburst galaxies,
could produce a much larger fraction  of the EGB.

\end{abstract}





\section{Introduction}
The extragalactic gamma-ray background (EGB) at GeV bands
was first detected by the satellite {\it Small Astronomy Satellite-2}
\citep[{\it SAS}-2;][]{fichtel75}, and its  spectrum was measured with good accuracy
by {\it Fermi} \citep[also called isotropic diffuse background;][]{abdo10a}.
Based on the first-year {\it Fermi} data, the EGB has been found to be
consistent with a featureless power law with a photon index of $\sim$2.4
in the 0.2--100\,GeV energy range and an integrated flux (E$\geq$100\,MeV)
of 1.03$\times10^{-5}$\,photon cm$^{-2}$ s$^{-1}$ sr$^{-1}$. The integrated
flux with E$\geq$1\,GeV is 4$\times10^{-7}$\,photon cm$^{-2}$ s$^{-1}$ sr$^{-1}$
\citep[see][]{abdo10b}.

The origin of the EGB is one of the fundamental unsolved problems in
astrophysics \citep[see][for a review]{k08}. The EGB could originate
from either truly diffuse processes or from unresolved
point sources. Truly diffuse emission can arise from some processes such as
the annihilation of dark matter \citep{ahn07,c10,b10},
the emission of high energy particles accelerated by intergalactic shocks which
are produced during large scale structure formation \citep{gb03} etc.

Blazars, including BL Lac objects, flat spectrum radio quasars, or
unidentified flat spectrum radio sources,
represent the most numerous population detected by the Energetic Gamma Ray
Experiment Telescope (EGRET) on the {\it Compton Gamma Ray
Observatory} \citep{h99} and by the {\it Fermi}-LAT \citep{abdo10d,A11}.
Therefore, blazars undetected by the EGRET or
{\it Fermi}-LAT are the most likely candidates for the emission of the EGB.
Many authors have studied the luminosity function of blazars and showed
that the contribution of undetected blazars to the EGB could be in the range
from 20\,\% to 100\,\% \citep{s96,n06,d07,c08,kn08,i09,L11,MH11,S11}.
However, \citet{abdo10b} built a source count distribution at GeV bands and yielded that
blazars undetected by the LAT can contribute about 23\,\% of the
EGB. They ruled out blazars producing a large fraction of the EGB.

Non-blazar radio loud active galactic nuclei (AGNs)
can also contribute a  fraction to the EGB \citep{b09,I11},
since  some of them have been found to be
$\gamma$-ray sources \citep{abdo10f} and they outnumber blazars.

Starburst galaxies, which have intense star-formation, are expected to have high
supernova (SN) rates. SN remnants are believed to accelerate primary cosmic rays (CRs)
protons and electrons. The high SN rates in starburst galaxies imply high CR emissivity.
When high energy CR protons collide with interstellar
medium (ISM) nucleons, they create pions decaying
into secondary electrons and positrons, $\gamma$-rays, and neutrinos.
\citet{A12} have found  quasi-linear scaling relations between $\gamma$-ray
luminosity and both radio continuum luminosity and total infrared luminosity.
To data, several starburst galaxies have been detected by the {\it Fermi}-LAT \citep{abdo10e, abdo11,A12}.
Therefore, starbursts can also contribute a fraction of the EGB \citep{t07,bs09,LT11,MR11}.

Nevertheless, normal galaxies \citep{bs09,F10} or radio-quiet AGNs \citep{I08, i09}
can also contribute a fraction of the EGB.

We introduced a new method of image stacking to directly study the undetected
but possible $\gamma$-ray point sources \citep{Z11}.   Applying the method to
the Australia Telescope 20 GHz Survey \citep[AT20G;][]{murphy10} sources
undetected by the LAT, we found that these sources contribute
about 10.5\,\%  of the EGB in the 1--3\,GeV energy ranges.

In this paper, we continually use the method to study the
origin of the EGB, and find that the FIRST sources undetected
by the LAT can contribute about (56$\pm$6)\,\% of the EGB.
We also discuss the implications.

\section{FIRST}


The Faint Images of the Radio Sky at Twenty centimeters
\citep[FIRST;][]{B95} program is using the Very Large Array
(VLA) to produce a map of the 20~cm (1.4~GHz) sky with a
beam size of 5.$''$4 and an rms sensitivity of about 0.15~mJy
beam$^{-1}$. The accuracy of source position is better than 1$''$.
The survey has covered an area of about 9,055
deg$^2$ in the north Galactic cap and a smaller area along the
celestial equator, corresponding to the sky regions observed
by the Sloan Digital Sky Survey (SDSS).
At the 1~mJy source detection threshold, the
source surface density is $\sim$ 90 deg$^{-2}$, and the
catalog includes $\sim8\times10^5$ sources\footnote{It is available online
http://sundog.stsci.edu/first/catalogs/readme\_08jul16.html}, in which
about 30\% of them have counterparts in the SDSS \citep{I02}.

The FIRST catalog lists two types of 20 cm continuum
flux density, e.g. the peak value, $F_{\rm peak}$, and the integrated flux
density, $F_{\rm int}$. These measurements are derived from
two-dimensional Gaussian fitting for each source, where the source
maps are generated from the co-added images from 12
pointings.

\citet{I02} found that about 30\,\% of FIRST sources was positional association within
1.$''$5 with an SDSS source in 1230 deg$^2$ of sky.
The majority (83\,\%) of the FIRST sources identified with an SDSS source brighter than
$r^*$ = 21 are optically resolved. About 30\,\% of them are radio galaxies
in which emission-line ratios indicate  AGNs; the others
are starburst galaxies. Nearly all optically unresolved
radio sources have nonstellar colors, indicating quasars.
Because there is no significant difference in the radio properties
between FIRST sources with and without optical identifications,
the majority of unmatched FIRST sources could be too
optically faint to be detected in SDSS images, and the fractions of
quasars and galaxies are roughly the same for the two subsamples.

Because two subsamples of the FIRST catalog are  $\gamma$-ray emitters,
this catalog maybe a good tracer for the undetected $\gamma$-ray point sources.




\section{Method}
For a sample of possible $\gamma$-ray point sources undetected by the
{\it Fermi} due to their faint fluxes, soft spectra \citep{abdo10c}
or the source confusion \citep{S11},
we can stack a large number of them to
improve the statistics  \citep{white07, ando10}.
We have used the Maximum Likelihood (ML) method to derive the fluxes
of the stacked point sources \citep{Z11}. But for a very large sample,
this method is very time consuming. Therefore, in this paper,
we use a  rough but simple ML method to derive the fluxes.

The photons\footnote{http://fermi.gsfc.nasa.gov/cgi-bin/ssc/LAT/LATDataQuery.cgi}
we used in our analysis are taken during the period of 2008 August
4 (15:43 UTC) -- 2011 October 20 (23:33 UTC), about three years.
During most of this time, {\it Fermi} was operated in sky-scanning survey mode
(viewing direction rocking north and south of the zenith on alternate orbits).
Time intervals flagged as `bad' (a very small fraction) and
the period of the rocking angle of the LAT greater than 52$^\circ$
was excluded.
Only the photons in the 1--100\,GeV energy range with
small 68\,\% containment radius (better than 1$^\circ$)
and little confusion \citep[see][]{at09,abdo09} are used.
These photons are also needed to satisfy the standard low-background event
selection (termed ``Source'' class events) corresponding to the
P7V6 instrument response functions in the present analysis.
The effect of Earth albedo backgrounds was
greatly reduced by removing photons coming from zenith angles $<100^\circ$.
In this procedure, the tools of {\it gtselect} and {\it gtmktime}
\footnote{These and other tools we used in next are parts of
most recent {\it Fermi}-LAT Science Tools, version v9r23p1,
which are accessible at  http://fermi.gsfc.nasa.gov/ssc/data/analysis/scitools/overview.html}
are used.

Stacking the images of the sources, we collect all photons that are
at most  $1^\circ$ away from any source of our sample and then record their
angular distance  ($\theta_i$, in units of deg) between the photon and the source.
The $1^\circ$ is enough despite some photons attributed to the central stacked
point sources are not in the stacked image.
Due to the faintness of the stacked
point sources,  in the outer of the stacked image,
the signal of the stacked point sources is very weak.

In order to minimize the influence of strong point sources,  the
sources away from any second  {\it Fermi}-LAT catalog \citep[2FGL;][]{abdo11}
sources  less $2^\circ$  ($3^\circ$ for 2GFL sources with fluxes larger than
$10^{-9}$~photon cm$^{-2}$ s$^{-1}$ in the 1--100\,GeV energy range)
are not used.
In order to minimize the influence of strong emission from the Galactic plane,
only the sources locating at high Galactic latitudes, e.g. $|b|>15^\circ$, are used.

There are a small number of spurious sources that are sidelobes of nearby bright sources.
In recent catalog, \citet{B03} uesd P(S) to indicate the probability of
that the source is spurious  (most commonly because it is a sidelobe
of a nearby bright source). The mean value of P(S)
is 0.088, which indicates that about 8.8\,\% of the sources are sidelobes.
They are concentrated around bright sources \citep[see][for detail]{W97,white13},
and should be removed first. In Figure 1 we show the number density of all
FIRST sources and the sources having $P(S)\le 0.1$
around the FIRST sources with $F_{\rm peak}\ge$100~mJy, respectively.
A strict P(S) cutting, such as P(S)$\le$0.1, is not appropriate.
We randomly remove some sources with the probability of their P(S)  except the sources
with  P(S)=0.014 (e.g., the minimum value of P(S), for about 72\,\% of all sources).
This procedure does not try to remove sidelobes as much as possible, but to
make the distribution of the rest sources around the bright sources as uniform as possible.
The number density of the rest sources around the FIRST sources
with $F_{\rm peak}\ge$100~mJy is also shown in Figure 1. The distribution
is still not uniform enough, we will investigate the effect of uniform distribution in the following section.

Many sources in the FIRST catalog are not independent objects but are
components of a single object with complex morphology \citep{W97}.
In Figure 2, we show the mean source number distribution of other
FIRST sources around  one FIRST source. It is found that
the possibility of finding another FIRST source near a FIRST source,
especially closer than $\sim50''$, is larger than one expected from
a random source distribution. Therefore, when a group sources fall within 50$''$
of their nearest neighbors,  we treat them as a single object.
Nearly 30\,\% of the FIRST sources belongs
to such sources. The cutoff of 50$''$ is a trade-off between
the completeness and contamination of the sample. For a larger cutoff, more complex sources
are merged, but more independent sources are also merged.
The use of 50$''$ cutoff will produce small uncertainty introduced by the misclassified sources,
and will be discussed in the following section.

The photon number density profile of the stacked image of FIRST sources with
$F_{\rm peak}\ge$1~mJy is shown in Figure 3. It is shown that
the center of the stacked image has higher photon number density
and the stacked point source has nonzero flux.

We use $\sigma_e(\theta)$ to further prove the presence of the point source,
which is the confidence level of stacked image representing photons within
the radius of $\theta$ more than the expectation in the uniform density. It is defined as
\begin{equation}
\sigma_e(\theta)=\frac{N(\theta)-\frac{N}{\Theta^2}\theta^2}{\sqrt{\frac{N}{\Theta^2}\theta^2}},
\end{equation}
where $N(\theta)$ is the number of photons with  $\theta_i<\theta$, $N$ is the number of
photons with  $\theta_i<\Theta$, here $\Theta=1^\circ$. The $\sigma_e(\theta)$
is presented in Figure  4.  The excess of
photons is very obvious within $\theta\sim0.3^{\circ}$.

we apply the maximum likelihood (ML) method to derive the flux
of the stacked point source.
The likelihood is the probability of the observed data for a
specific model. For our case, it is defined as
\begin{equation}
L=\prod_{i=1}^{33}p_i,
\end{equation}
where
\begin{equation}
p_i={\frac{1}{\sqrt{2\pi N_i}}\exp\left[-\frac{(N_i-M_i)^2}{2N_i}\right]}
\end{equation}
is the normal probability of observing $M_i$ counts in $i$-th
bin when the number of counts predicted by model is $N_i$.
The logarithm of the likelihood is conveniently
calculated:
\begin{equation}
\ln L=\sum_{i=1}^{33}\left[-0.5\ln(2\pi)-0.5\ln N_i-\frac{(N_i-M_i)^2}{2N_i}\right].
\end{equation}
For simplicity, in our model there are
only two components, e.g., the central stacked point source and  diffuse
background source, and
\begin{equation}
N_i=N_sn_{si}+N_bn_{bi},
\end{equation}
\begin{equation}
N=N_s+N_b,
\end{equation}
where $N_s$ and $N_b$ are the photon numbers attributed to the central stacked point source
and the diffuse background source, respectively.
$n_{si}$ and $n_{bi}$ are the normalized
photon number distributions of two class sources.

In order to determine $n_{bi}$, we  stacked $10^8$
imaginary sources which is isotropically distributed on the sky.
In fact, only $\sim4 \times 10^7$ sources far away
strong sources are ultimately stacked. Because
the sources we stacked here are not true $\gamma$-ray point
sources, the stacked image only contains diffuse backgrounds.
The photon number density profile of the stacked image is shown
in Figure 5. It is found that the diffuse background is
not uniform and has higher density in the outer of the
stacked image, implying that the profile is affected
by the 2FGL sources near the stacked image.
In Figure 6 we show the photon number density profile of the stacked image
which is extended to 2$^\circ$, indicating that the effect of the nearby point sources is much
obvious.

In order to determine $n_{si}$, we simulate a source with a flux of
$5 \times 10^{-6}$~photon cm$^{-2}$ s$^{-1}$  in the 1--100\,GeV energy range
and the coordinate of (RA, DEC) = (190$^\circ$, 30$^\circ$)\footnote{
This point has similar exposure to the stacked point sources, see Figure 7.}.
It has a power law spectrum and
the photon index is 2.4, which is a typical value of the detected sources.
The tool {\it gtobssim} has been used in this procedure.
The photon number density profile of this source is shown in Figure 8.

We use the likelihood ratio to test the hypothesis. The point-source
``test statistic'' (TS) is defined as
\begin{equation}
{\rm TS}=-2(\ln L_0-\ln L_1)
\end{equation}
where $L_1$ and $L_0$ are the likelihood with and without point source.
The TS of each source is related to the probability of that
such an excess is obtained from background fluctuations
alone. The probability distribution in such a situation is not known precisely \citep{P02}.
However, we only consider positive fluctuations, in which each
fitting involves one degree of freedom, and
according to Wilks's theorem \citep{W38} the probability
with at least TS is close to that of the $\chi^2/2$
distribution with one degree of freedom. Therefore
the detected significance of a point source is approximately
$\sqrt{\rm TS}\sigma$ \citep[see][]{m96}.

\section{Result and Discussion}
For the stacked point source of the FIRST sources with $F_{\rm peak}\ge$1~mJy,
the estimated $N_s$ is $2.51 \times 10^4$, in which its TS is 43.0,
corresponding to a significance of $\sim6.7\sigma$.
The $\chi^2$ over degree of freedom (dof) is $\chi^2/{\rm dof}=30.7/31$,
indicating that the model gives a satisfactory fit to the data.

$N_s$ is proportional to the flux and exposure of the
stacked point source. The exposure is the mean of the exposures of
all stacked sources. But the calculation of the
exposures\footnote{Calculated using the tool {\it gtpsf}~ in {\it Fermi}-LAT Science Tools.} for
all stacked sources is very time consuming, and the  exposure of all sky is relatively uniform,
owing to the large field-of-view and the rocking-scanning pattern of the sky survey \citep{abdo11},
we use the mean exposure of about 1\,\% randomly selected FIRST
sources to represent them. The exposures of the stacked point source
and the simulated source are shown in Figure 7, which are roughly the same. The simulated source has a flux of
$5\times10 ^{-6}$~photon cm$^{-2}$ s$^{-1}$  and $\sim4.1 \times 10^5$
photons within $1^\circ$. The stacked point source has $\sim2.51 \times 10^4$ photons
within $1^\circ$, corresponding to a flux of $\sim3.09 \times 10^{-7}$~photon
 cm$^{-2}$ s$^{-1}$. The FIRST covers an area of about 9055 deg$^2$, in which
321556 out of 640919 sources are stacked, indicating that the stacked point sources cover an area
( i.e. the area covered by the FIRST sample excluding
all holes produced in the {\it Fermi} map by excluding photons in the vicinity
of  2FGL sources.) of about
 4543 deg$^2$. In this sky region the total flux of the EGB is $\sim5.7 \times 10^{-7}$~photon
 cm$^{-2}$ s$^{-1}$, implying that these sources can contribute $\sim56\,\%$ of the EGB.

A decrease of 0.5 from its maximum value in $\ln L$ corresponds
to 68\,\% confidence (1$\,\sigma$) region for the parameter \citep[see][]{m96}.
We use this variance to estimate the error of the flux
and find that 1$\,\sigma$ error is $0.33\times 10^{-7}$~photon cm$^{-2}$ s$^{-1}$,
implying that the FIRST sources contribute about (56$\pm$ 6)\,\% of the EGB.

The effective area is an important factor of uncertainties. The
current estimate of the remaining systematic uncertainty is 10\,\%
at 100 MeV, decreasing to 5\,\% at 560 MeV and increasing to
10\,\% at 10 GeV and above \citep{A12b}. This uncertainty is uniformly applied
to all sources \citep{abdo11}. The error of the fraction of FIRST sources
contributed to the EGB is much smaller.

In order to verify the effectiveness and accuracy of our method,
we do the Monte Carlo simulation using the tool {\it gtobssim}.
The simulating time is $\sim10^8$\,s, equaling the time
of real data we used. We simulate $7.8\times10^5$ point sources
with fluxes of $10^{-12}$\,photon cm$^{-2}$ s$^{-1}$
in the 1--100\,GeV energy range. The flux distribution does not
affect our results, which only depend on the total flux.
They have  power law spectra, in which the photon index of each source is drawn from a Gaussian
distribution with a mean of 2.40 and $1\,\sigma$ width of 0.24,
which is the same with the intrinsic  distribution of photon indices \citep{abdo10b}.
The coordinates of each source are randomly drawn to produce an
isotropic distribution on the sky between right ascensions of 95$^\circ$
and 270$^\circ$, declinations of -10$^\circ$ and +70$^\circ$, which
covers the main area observed by the FIRST survey. The simulated sources have similar
source surface density with the FIRST sources, in which some sources are
randomly removed according to the probability of their P(S)  nd sources separated by less
than 50$''$ are merged to a single source.

The simulated photons are isotropic in this area except edge part.
We model the Galactic diffuse background using the models (gal\_2yearp7v6\_v0.fits)
recommended by the LAT team, although it only contributes an uniform
background to the stacked image. The EGB is not modeled, because the photons from
point sources are actually regarded as the EGB.

To avoid the edge effect, we only stack the sources with the coordinates
between right ascensions of 100$^\circ$ and 265$^\circ$, declinations
of -5$^\circ$ and +65$^\circ$. Only about 2.2$\times10^5$ of randomly
selected simulated sources are stacked to obtain the
similar flux with the stacked source of the FIRST sources. Other $\sim1\times10^5$
random positions are also stacked to obtain the similar photon number
density with the stacked image of the FIRST sources.
The photon number density profile of the stacked image
is shown in Figure 9. Then we use the ML method to obtain 
the mean flux, which is 1.08$\times10^{-12}$\,photon cm$^{-2}$ s$^{-1}$.
Its TS is 41.9, and $\chi^2$ is 31.7. The $1\,\sigma$ error is 1.5$\times10^{-13}$
\,photon cm$^{-2}$ s$^{-1}$. Therefore, the derived mean flux is equal to the
input flux within $1\,\sigma$ error, indicating that our flux estimation is correct and reliable.

The stacked image contains some regions which are not independent, leading to
most photons being counted more than once. If a photon comes
from other point source rather than the center source, it will be
regarded as a part of the backgrounds and does not affect our results, as proved by the simulation.

In order to investigate the effect of the complex source merging and
the sidelobe removing on our results, we use different $\theta_{cut}$ to merger
the complex sources. The result, shown in Figure 10, presents that the fraction
decreases when the $\theta_{cut}$ increases. This is due to two factors: (1) many
sidelobes and complex sources are merged; (2) many independent
sources are wrongly merged.

For more detail, we also merger the simulated sources
using various $\theta_{cut}$ to calculate the total flux of the samples,
in which the Galactic diffuse background is not added. The result is shown in Figure 10.
The slope of the first curve is obviously steeper than that of the second curve when $\theta_{cut} \lesssim 100''$,
but the slopes of two curves are nearly the same when $\theta_{cut} \gtrsim 100''$. This indicates that
the effect of sidelobe and complex source merging is obvious only when $\theta_{cut} \lesssim 100''$.
However, when $\theta_{cut} = 50''$,
the effect of  the complex source merging and the sidelobes\ removing is small, and only makes
the fraction to be overestimated by about 2\,\%. The wrong merging
of independent sources makes the fraction to be also underestimated by about 2\,\%.
Therefore, the cutoff of 50$''$ is appropriate and only introduces a small uncertainty.

Considering that the FIRST is
a flux limited sample of radio sources and incomplete at the faint limit \citep{W97},
we consider that when the sample is more complete or its radio flux limit further decreases,
the contribution of the FIRST sources to the EGB should increases, but the exact fraction is not clear
because it depends on the shape of the radio source count distribution below
the FIRST flux limit and the correlation between the source radio and $\gamma$-ray luminosity.
Nevertheless, normal galaxies \citep{bs09,F10}
or radio-quiet AGNs \citep{I08, i09}, which cannot be well traced by the FIRST survey
(only a few fraction of those sources can be included in FIRST),
can also contribute a fraction of the EGB, we think that the point sources can
contribute most of the EGB.

\section{Conclusions}
In the paper, we use the method of image stacking to study the
origin of the EGB, and find that the FIRST sources undetected by LAT can contribute about
(56$\pm$6)\,\% of the EGB. Considering the flux limit and incompleteness
of the sample at the faint limit,
we think that most of the EGB is distributed by point sources
which are not resolved by LAT because of source confusion and weak flux.
The main contributors of the EGB maybe blazars, non-blazars AGNs and
starburst galaxies.
But it is difficult to derive the exact fraction of each population
contributing to the EGB using our method alone.

\acknowledgments
\section*{Acknowledgments}
We thank the LAT team, AT20G team and FIRST team providing the data on the website.
We acknowledge the financial supports from the National
Basic Research Program of China (973 Program 2009CB824800),the National Natural
Science Foundation of China 11133006, 11163006, 11173054,  and the
Policy Research Program of Chinese Academy of Sciences (KJCX2-YW-T24).

\clearpage

\begin{figure}
\plotone{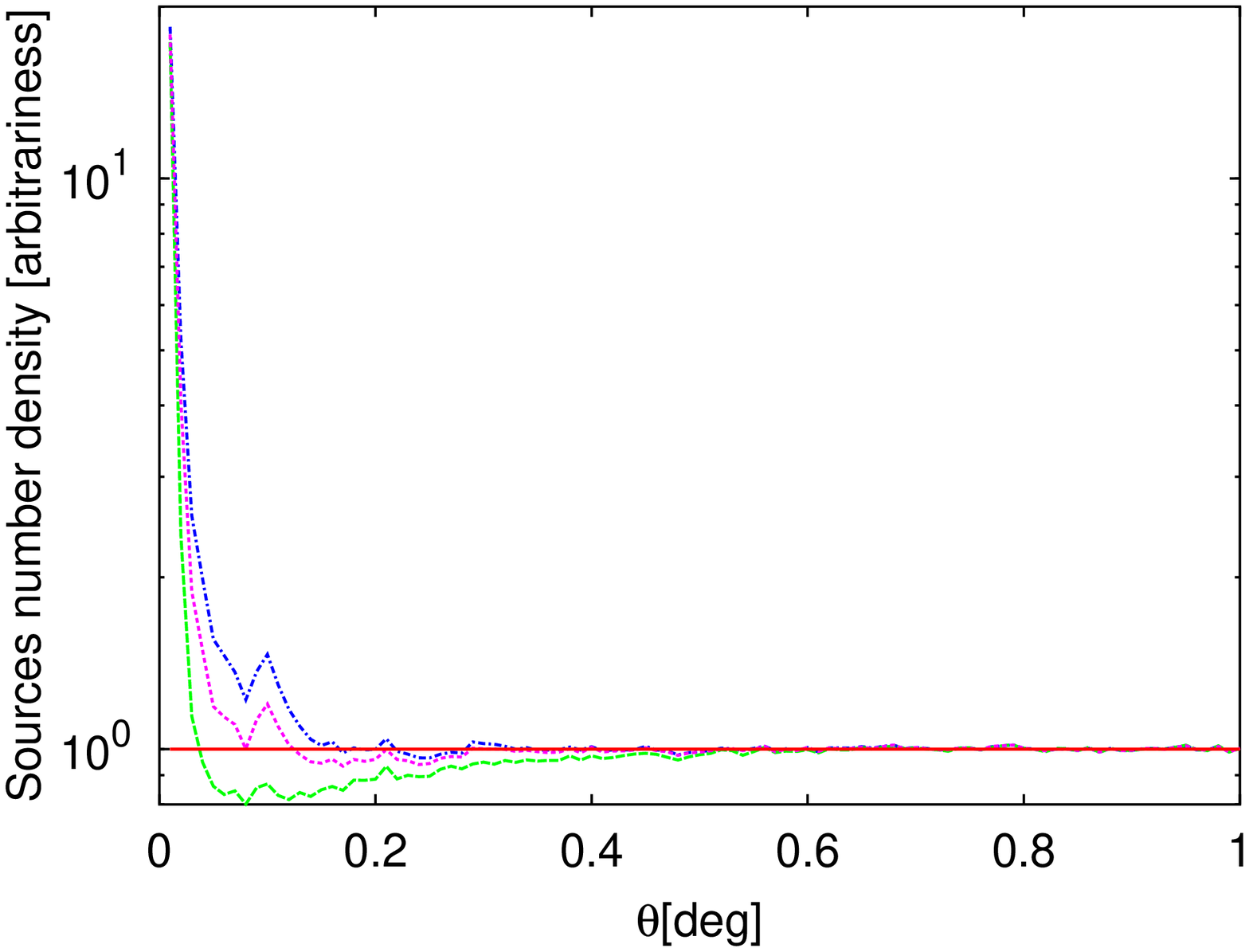}
\caption{The  normalized source number density profiles  around the FIRST sources
with $F_{\rm peak}\ge$100~mJy, for all sources by dot-dashed (blue) line, for the sources randomly removed with the probability of
their P(S) by dotted (magenta) line, and for the sources with
P(S)$<$0.1 by dashed (green) line. The solid (red) line represents the uniform density profile.}
\end{figure}

\begin{figure}
\plotone{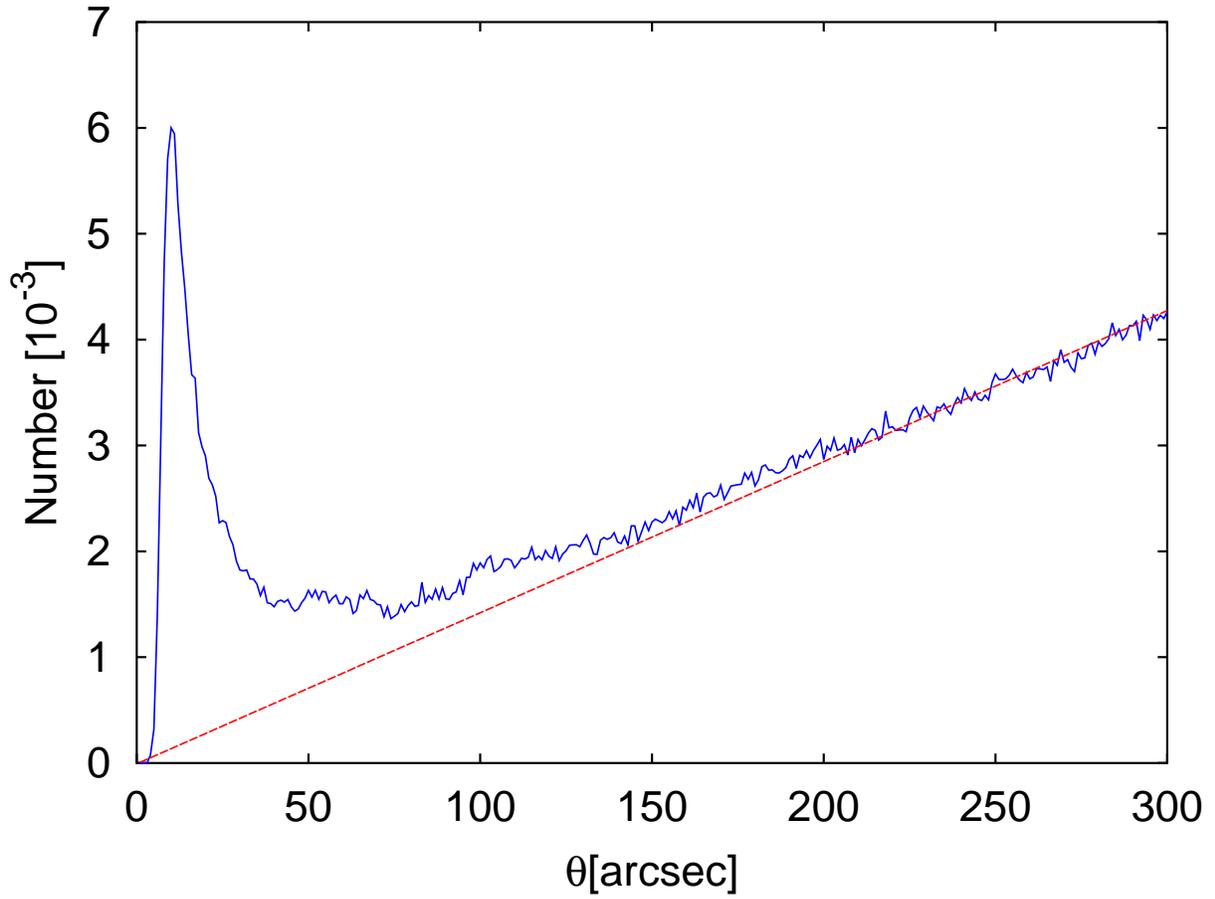}
\caption{The mean number distribution of FIRST sources around one FIRST source. The dashed (red)
line represents an estimate of the chance coincidence rate.}
\end{figure}

\begin{figure}
\plotone{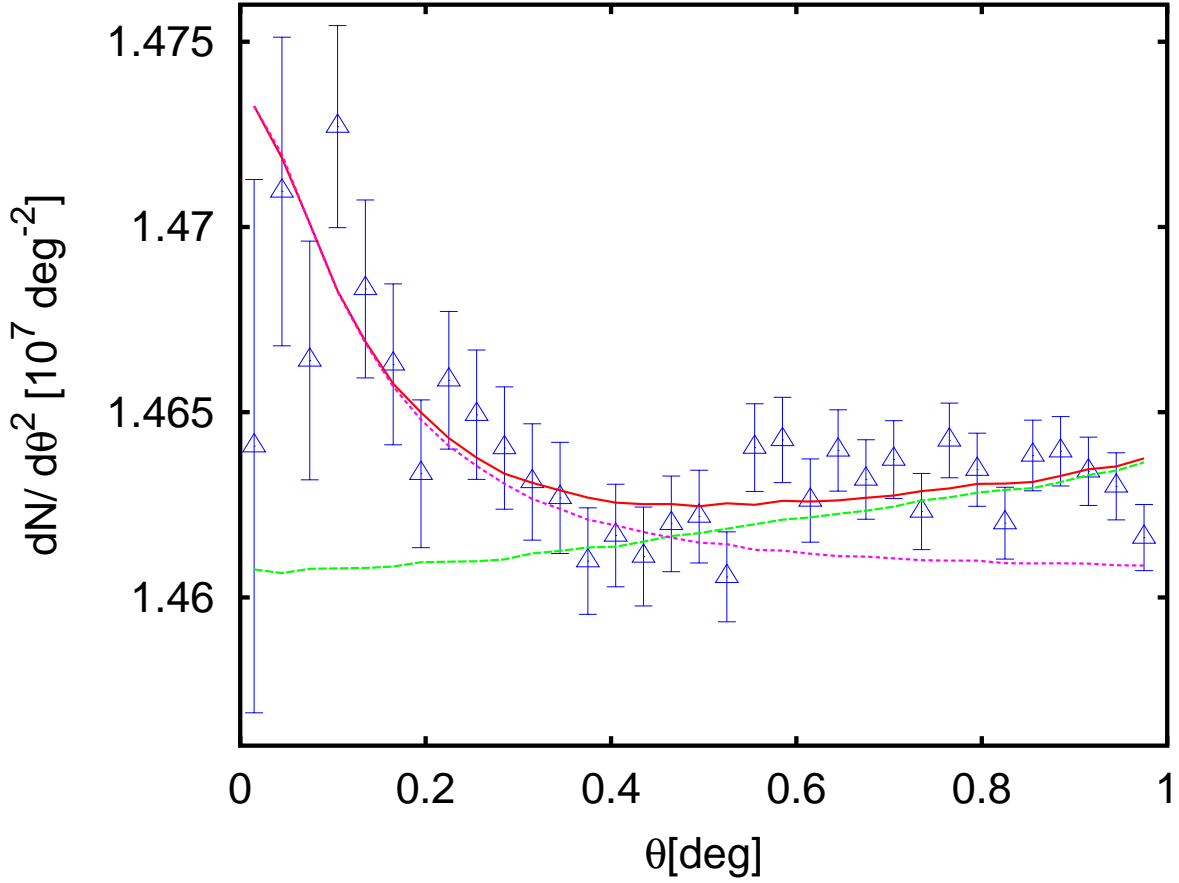}
\caption{The photon number density profile of the stacked FIRST sources with
$F_{\rm peak}\ge$1~mJy. The dashed (green) line
represents the diffuse background, while the dotted (magenta) line represents the point source 
added by an arbitrary factor. The solid (red) line represents the best-fit model.
The $\chi^2$ over degree of freedom (dof) is $\chi^2/{\rm dof}=30.7/31$,
indicating that the model gives a satisfactory fit to the data.
The TS is 43.0, corresponding to a significance of $\sim6.7\sigma$.}
\end{figure}

\begin{figure}
\plotone{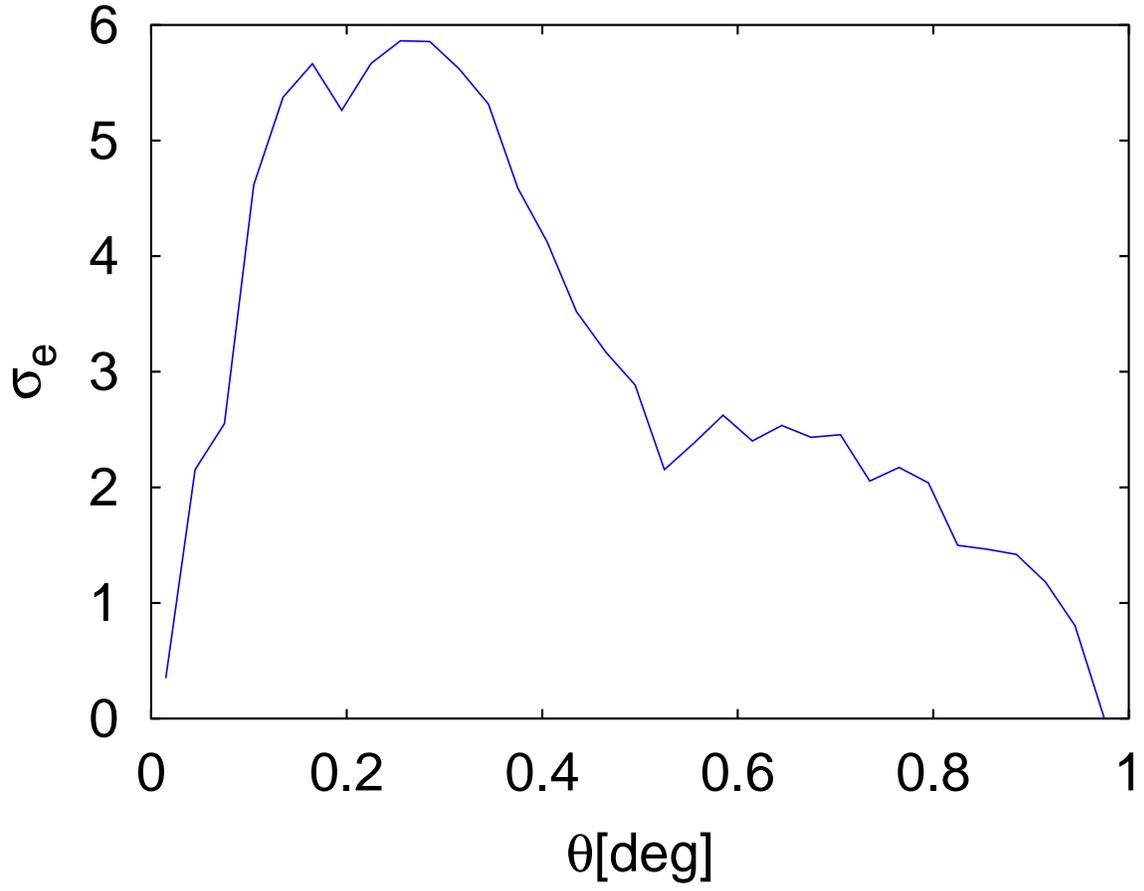}
\caption{$\sigma_e(\theta)$ for the stacked image of the  FIRST sources.}
\end{figure}

\begin{figure}
\plotone{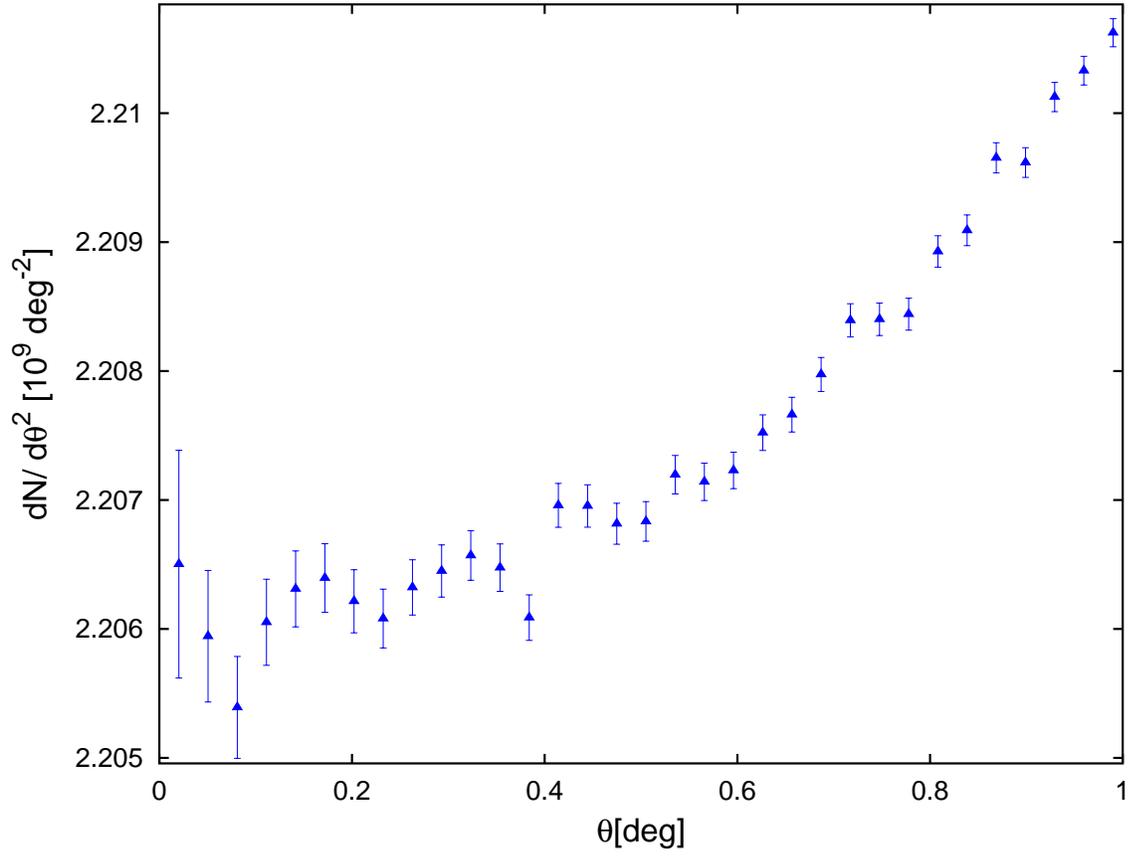}
\caption{The photon number density profile of the stacked imaginary sources.}
\end{figure}

\begin{figure}
\plotone{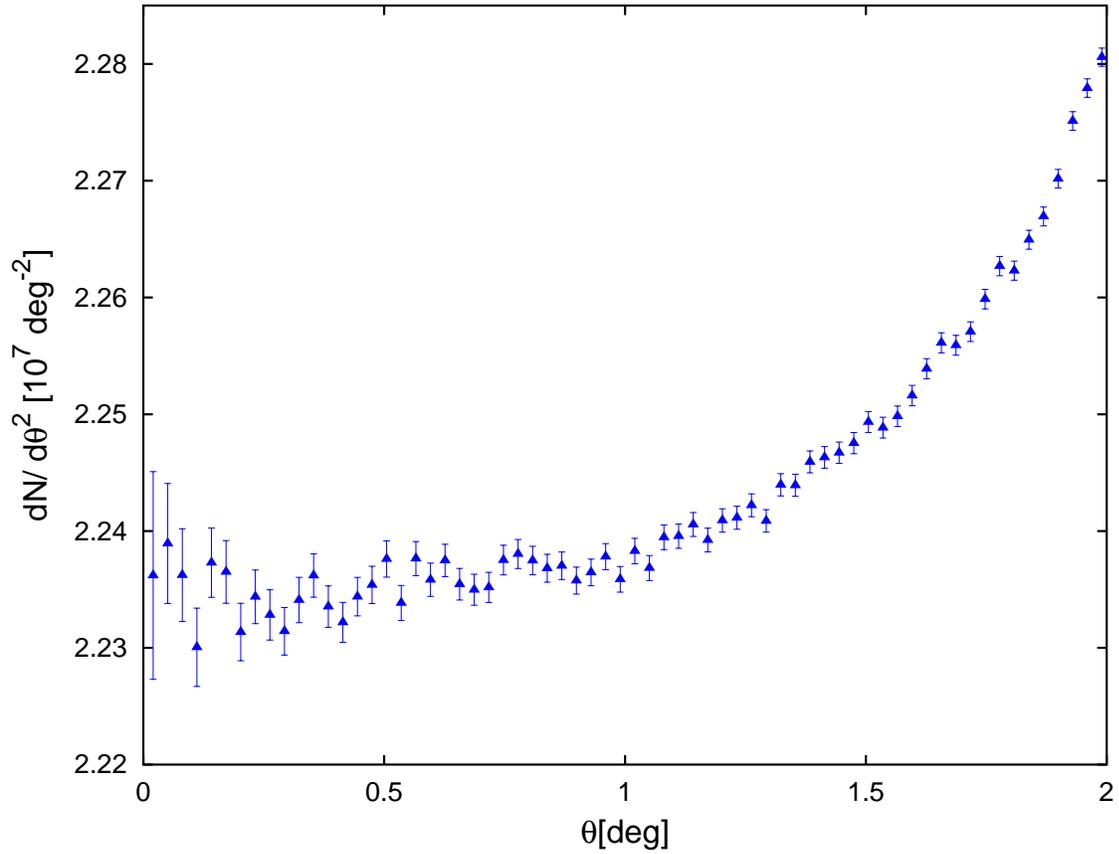}
\caption{Same as Figure 5, but angular range is extended to 2$^\circ$ and
a small number of imaginary sources is stacked.
In the figure, the effect of the nearby 2FGL sources is more obvious.}
\end{figure}

\begin{figure}
\plotone{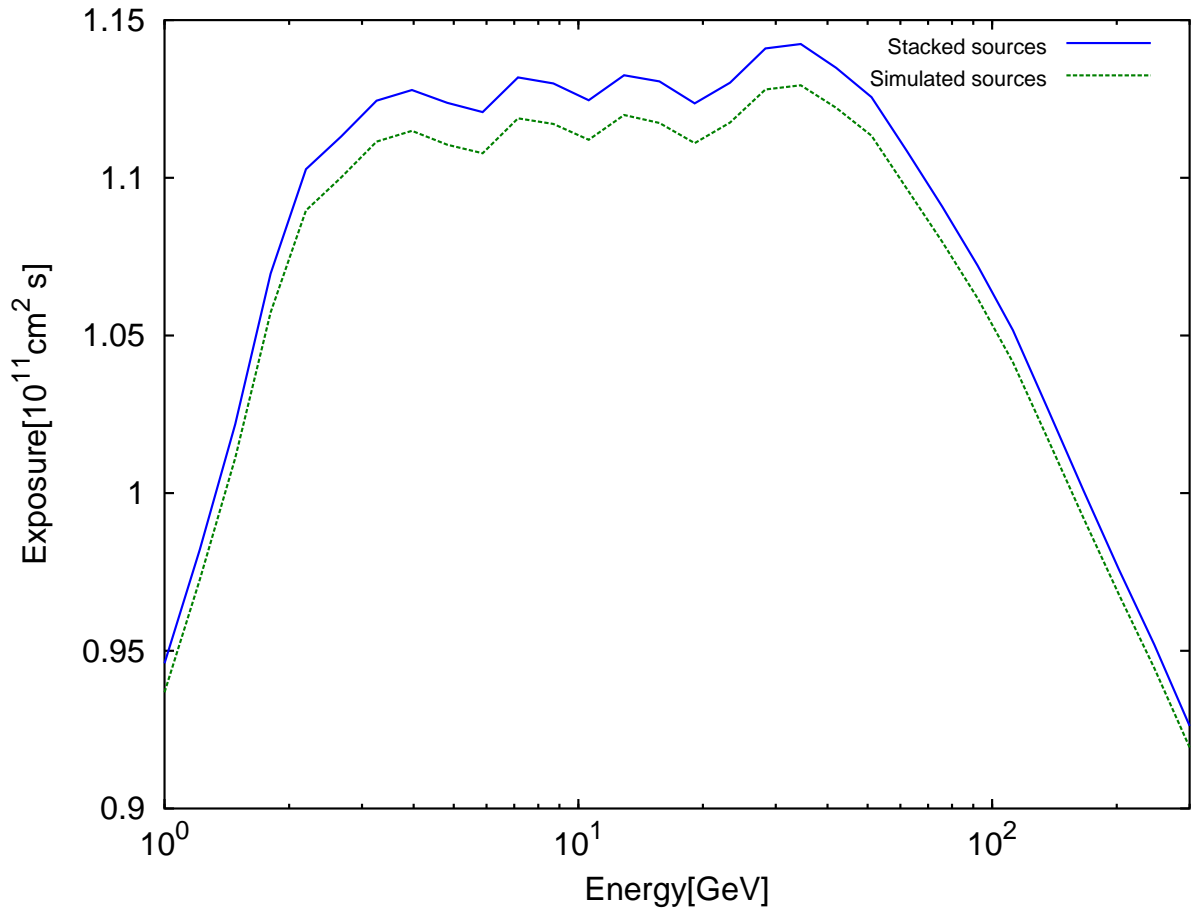}
\caption{The exposures of the stacked source and the simulated source with coordinates of
(RA, DEC) = (190$^\circ$, 30$^\circ$ ).}
\end{figure}

\begin{figure}
\plotone{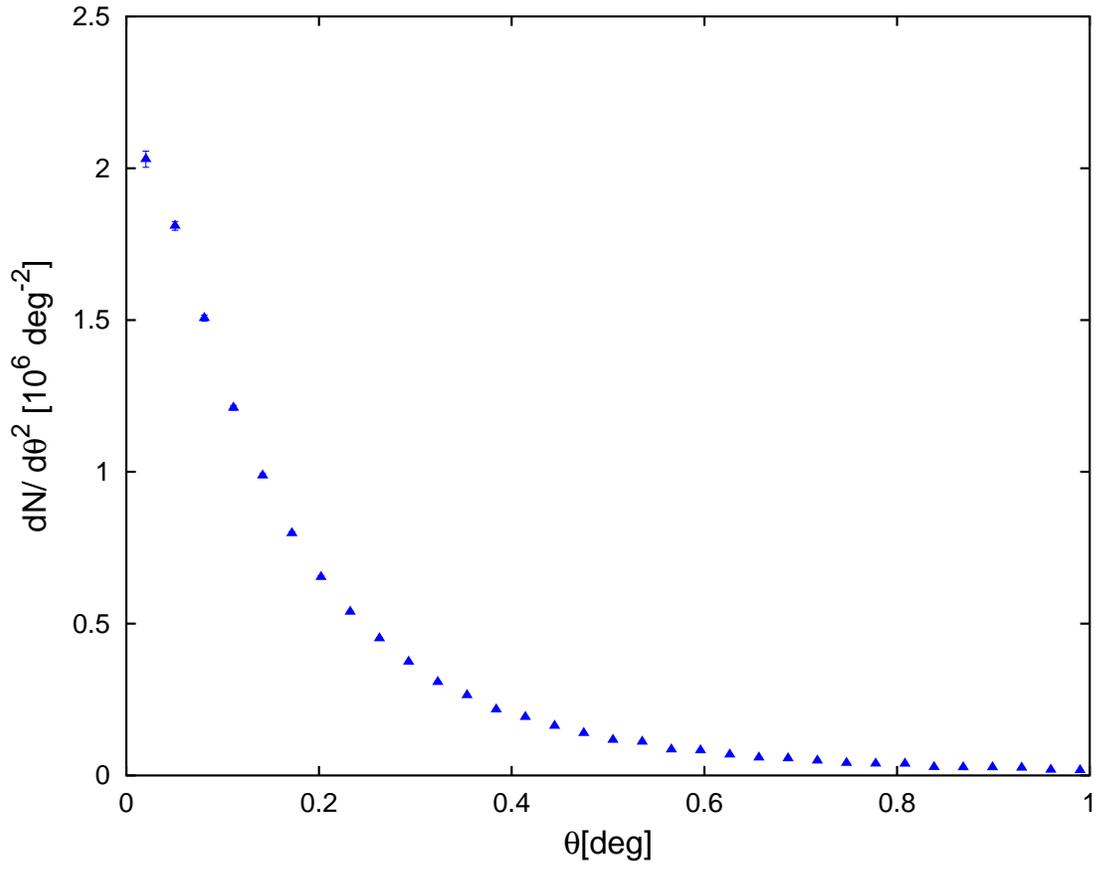}
\caption{The photon number density profile  of the simulated source with coordinates of
(RA, DEC) = (190$^\circ$, 30$^\circ$ ).}
\end{figure}

\begin{figure}
\plotone{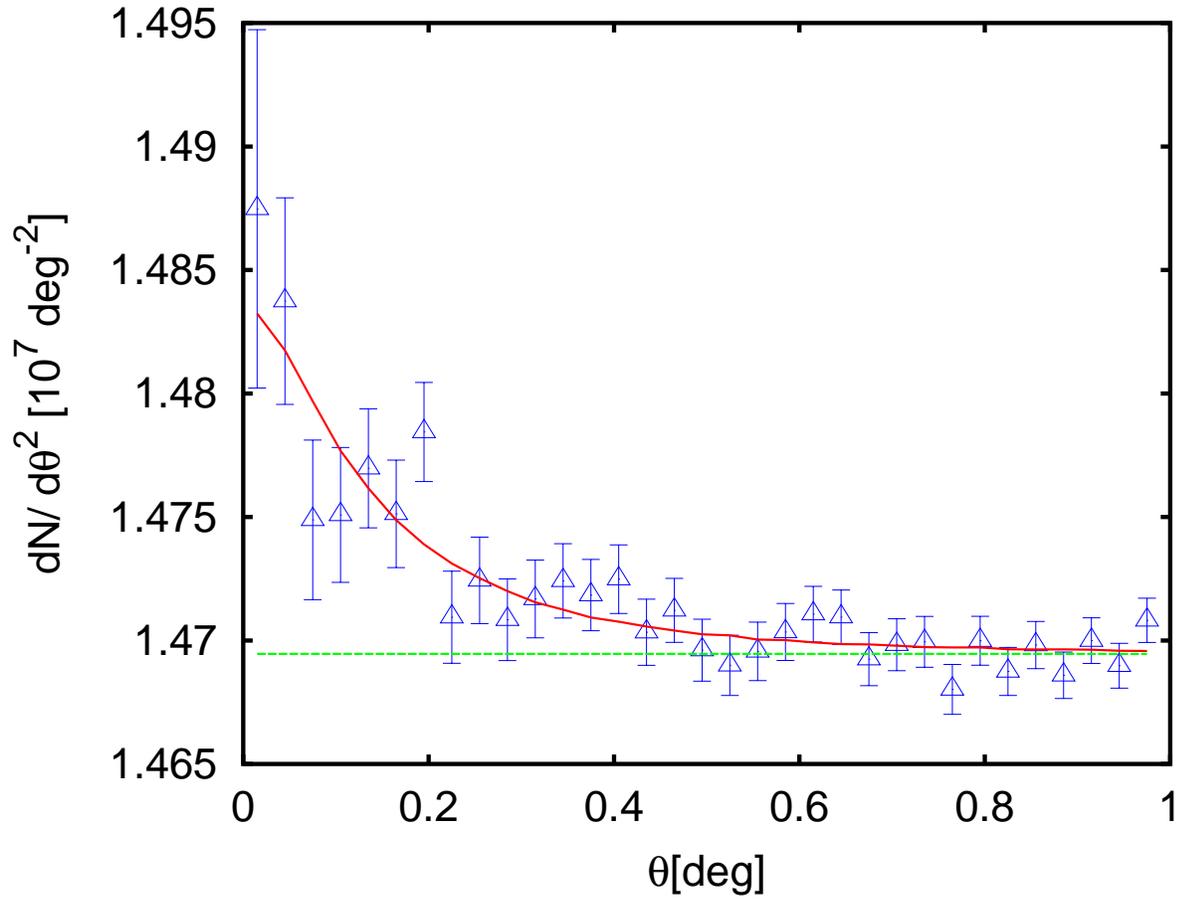}
\caption{The photon number density profile of the stacked image
using the simulated data. The dashed (green) line
represents the diffuse background which is uniform. The solid (red) line represents the best-fit model, in which
TS is 41.9, and $\chi^2$ is 31.7.}
\end{figure}

\begin{figure}
\plotone{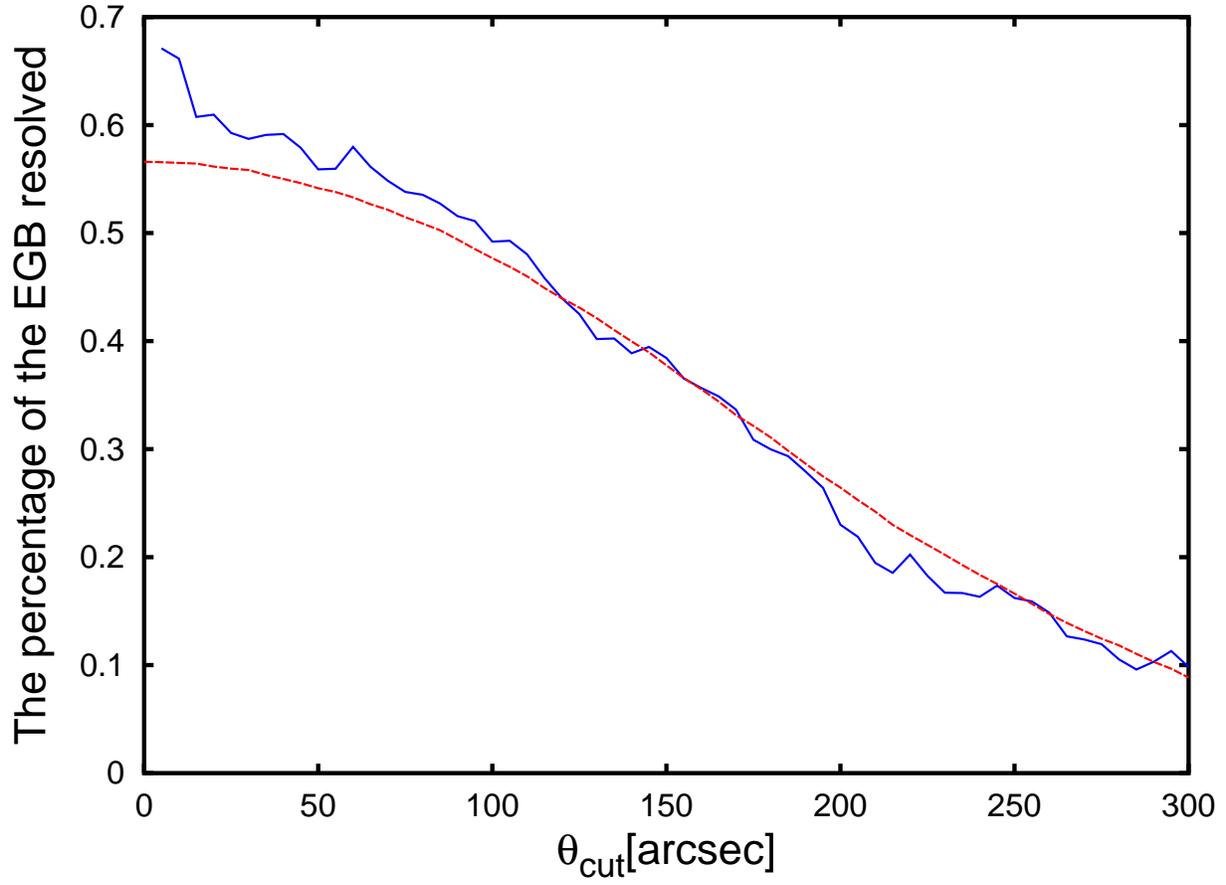}
\caption{The percentage of the EGB resolved as a function of angular cut in the FIRST catalog.
The dashed (red) line represent the expected dependence of the percentage on the angular cut for a random
source distribution, which is obtained from the simulation. For comparison,
two curves are normalized to have same value at  $\theta_{cut}=120''$.}
\end{figure}

\end{document}